\documentclass[aps,pra,twocolumn,longbibliography]{revtex4-1}
\usepackage{amsmath}
\usepackage{amssymb}

\newcommand\beq{\begin{equation}}
\newcommand\eeq{\end{equation}}
\newcommand\bea{\begin{eqnarray}}
\newcommand\eea{\end{eqnarray}}
\newcommand\nn{\nonumber}

\newcommand\ro{\hat\rho}
\newcommand\Po{\hat P}

\newcommand\Ho{\hat H}
\newcommand\Lo{\hat L}
\newcommand\VoG{\hat{V}_\mathrm{G}}
\newcommand\dg{^\dagger}

\newcommand\eps{\epsilon}
\newcommand\der{\partial}
\newcommand\Io{\hat I}
\newcommand\tr{\mathsf{tr}}
\newcommand\Herm{\mathbb{H}}

\newcommand\Mean{\mathbb{M}}
\newcommand{\Schr}{Schr\"odinger}

\newcommand{\al}{\alpha}
\newcommand{\be}{\beta}
\newcommand{\si}{\sigma}
\newcommand{\la}{\lambda}
\newcommand{\ga}{\gamma}
\newcommand{\Loa}{\Lo^\al}

\newcommand\DQ{D^\mathrm{Q}}
\newcommand\DC{D_\mathrm{C}}
\newcommand\GCQ{G_\mathrm{CQ}}
\newcommand\GCQb{\overline{G}_\mathrm{CQ}}

\newcommand{\bD}{\mathcal{D}}

\newcommand\HCl{H_\mathrm{C}}
\newcommand\HoQ{\Ho_\mathrm{Q}}
\newcommand\HoCQ{\Ho_\mathrm{CQ}}
\newcommand\Phio{\hat{\phi}}
\newcommand\Lc{\mathcal{L}}
\newcommand\muo{\hat{\mu}}
\newcommand\wt{\widetilde{w}}
\newcommand\Phimf{\Phi_\mathrm{mf}}

\newcommand\enm{\eps^{nm}}

\newcommand\Dn{_{,n}}
\newcommand\Dm{_{,m}}

\newcommand\Ao{\hat A}
\newcommand\Bo{\hat B}

\newcommand\vo{\hat v}
\newcommand\rv{\mathbf{r}}
\newcommand\sv{\mathbf{s}}
\newcommand\HG{H_\mathrm{G}}
\newcommand\HoM{\Ho_\mathrm{M}}
\newcommand\Hc{\mathcal{H}}
\newcommand\HcG{\Hc_\mathrm{G}}
\newcommand\Hco{\hat{\Hc}}
\newcommand\HcoM{\Hco_\mathrm{M}}
\newcommand\Pc{\mathcal{P}}

\newcommand\sqrtg{\sqrt{g}}

\begin{document}
\title{The classical-quantum hybrid  canonical dynamics and its
difficulties with special and general relativity}
\author{Lajos Di\'osi}
\email{diosi.lajos@wigner.hu}
\homepage{www.wigner.hu/~diosi} 
\affiliation{Wigner Research Center for Physics, H-1525 Budapest 114 , P.O.Box 49, Hungary}
\affiliation{E\"otv\"os Lor\'and University, H-1117 Budapest, P\'azm\'any P\'eter stny. 1/A, Hungary}
\date{\today}

\begin{abstract}
We discuss the Hamiltonian hybrid coupling between a classical and a quantum subsystem. 
If applicable to classical gravity coupled to quantized matter,
this hybrid theory might realize a captivating `postquantum' alternative to full 
quantum-gravity. We summarize the  nonrelativistic hybrid dynamics in improved
formalism adequate to Hamiltonian systems. The mandatory decoherence and diffusion
terms become divergent in special and general relativistic extensions. 
It is not yet known if any renormalization method might  
reconcile Markovian decoherence and diffusion with relativity.
Postquantum gravity could previously only be realized in the
Newtonian approximation. We argue that pending problems of the recently proposed
general relativistic postquantum theory will not be solved if
Markovian diffusion/decoherence are truly incompatible with relativity.
\end{abstract}
\maketitle

\section{Introduction}\label{sec_Intr}
The dynamical coupling between a classical and a quantum subsystem
is of multiple interests, e.g.,  in mathematical physics, in
heuristic models, and particularly in foundations. If gravity
were fundamentally classical then its hybridized dynamics with
quantized matter would replace the mean-field (semiclassical) approximation 
\cite{moller1962,rosenfeld1963} and the famously inconclusive
versions of quantum-gravity. Such a captivating idea has been kept alive
from  an episodic suggestion  \cite{maddox1995,anderson1995} 
---based on incorrect nonrelativistic (NR) hybrid dynamics 
\cite{diosi1996Comment}---
through works by the present author and by others
\cite{diosi1989,diosi1990,diosi1995twoplanck,
          diosi1998coupling,
          diosi2011gravityrelated,
          tilloydiosi2016sourcing,tilloydiosi2017} until to culminate in
 \emph{postquantum} gravity of Oppenheim and co-workers
\cite{oppenheim2022constraints,layton2022healthier,
          oppenheim2023gravitationally,oppenheim2023postquantum}.

Parallel to the fundamental concept,  the underlying mathematical tool has been
researched persistently along important milestones 
\cite{aleksandrov1981,gerasimenko1982,boucher1988,
          diosi1995twoplanck,diosi1998coupling,diosi2000hybrid,
          diosi2000quantum,diosi2011gravityrelated,diosi2014hybrid,
          blanchard1993,blanchard1995,alicki2003,
          oppenheim2022two,oppenheim2023objective,
          diosi2023hybrid}.
The central technical issue that has been solved non-relativistically 
is the following.
Suppose the hybrid Hamiltonian contains in turn the classical Hamilton
function of the classical subsystem, the Hamilton operator of the
quantum subsystem, and the coupling between them:
\beq
\Ho(q,p)=\HCl(q,p)+\HoQ+\HoCQ(q,p).
\eeq
The evolution equation of the state-vector of the quantum subsystem is the
\Schr~ equation $i\hbar|\Psi\rangle/dt=\Ho(q,p)|\Psi\rangle$. The
\emph{backaction} of the quantum subsystem on the classical one is
non-trivial. Towards the solution of interest, we introduce
the hybrid state,
represented by the hybrid density $\ro(q,p)\geq0$
which is a combination of the density operator 
$\ro_\mathrm{Q}=\int\ro(q,p)dqdp$ of the quantum
subystem and the phase space density $\rho_\mathrm{C}(q,p)=\tr\ro(q,p)$ of the classical one.
Assume the following combination of the classical and quantum
dynamics  \cite{aleksandrov1981}: 
\bea\label{Ale}
\frac{d\ro(q,p)}{dt}&=&-\frac{i}{\hbar}[\Ho(q,p),\ro]+\Herm\{\Ho(q,p),\ro(q,p)\}\nn\\
                                     &\equiv&\{\Ho(q,p),\ro(q,p)\}_\mathrm{A}\nn\\
\eea
where $\{~,~\}$ stands for the Poisson bracket. The term $\Herm\{\Ho_{CQ}(q,p),\ro(q,p)\}$
represents the backaction, the symbol $\Herm$ means the Hermitian part. 
If it is zero, we get the standard classical and quantum dynamics 
separately for the two subsystems, as we should. 
But the seemingly plausible dynamics (\ref{Ale})
is not yet mathematically correct, it does not preserve the positivity
of $\ro(q,p)$. Additional decoherence and diffusion mechanisms are 
mandatory and they are subject of  trade-off: stronger decoherence 
allows for weaker diffusion and vice versa \cite{diosi1995twoplanck}. 
The ultimate general form of hybrid NR dynamics appeared in refs. 
\cite{oppenheim2022two,oppenheim2023objective,diosi2023hybrid,diosi2023erratum}.    

Instead of a master equation 
for $\ro(q,p)$,  stochastic differential equations
for the pure quantum state $\Po$ and the classical variables $(q,p)$
offer an equivalent alternative. As an analogy, remember for example
that the classical Fokker--Planck equation is equivalent to the
Langevin stochastic differential equation. In the hybrid case, 
the backaction is realized by time-continuous quantum measurement 
---\emph{monitoring}--- of the quantum subsystem and \emph{feedback} 
of the measured signal into the classical subsystem. 
The importance of this formalism is emphasized especially in refs.
\cite{diosi2023hybrid,tilloy2024general}.
Compared to the master equation of hybrid canonical coupling, the
modular monitoring-plus-feedback construction gives better intuition
as observed in ref. \cite{tilloy2024general}. 

Undoubtedly, hybrid coupling is not possible without compromises. 
For example, there are two fundamental issues with the semiclassical approximation:
fake action-at-a-distance and breakdown of Born's statistical interpretation 
(cf., e.g., notes \cite{diosi2016nonlinear} and refs. therein). 
These two issues would be unacceptable in a fundamental theory. 
But, it is crucial that they are NR effects 
absolutely unrelated to relativity or gravity, 
related only to the non-linearity of the semiclassical hybrid 
equations. It was shown that linearity can be maintained by assuming a well-defined 
minimum noise in the hybrid coupling \cite{diosi1995twoplanck}. We thus have a linear NR hybrid dynamics 
\cite{oppenheim2022two,oppenheim2023objective,diosi2023hybrid,diosi2023erratum}
with tractable compromise (noisiness) instead of the two fundamental defects
mentioned above. 
The question now is whether there is a relativistic extension of this hybrid dynamics.

Our goal is threefold:
a convenient introduction to the mathematics of
NR hybrid canonical dynamics, the assessment of its application in
postquantum gravity and a discussion if it could have surpassed
its old Newtonian `forerunner'. 
 
Section \ref{sec_nonrel} recapitulates state-of-the-art knowledge
of NR hybrid canonical dynamics. 
Section \ref{sec_rel} explains the locality condition of relativistic
invariance and the resulting divergences.
Section \ref{sec_specrel} tests
the special relativistic extension on the simplest example 
of hybrid coupling between a classical and a quantized scalar field.
Section \ref{sec_genrel} revisits the effort towards general relativistic
postquantum gravity, extending  the NR hybrid dynamics for general relativity.
Section \ref{sec_DP} recapitulates the NR
`forerunner' of postquantum general relativity.
Final remarks and our conclusion are given by sec. \ref{sec_remarks}. 

\section{The nonrelativistic canonical hybrid dynamics}\label{sec_nonrel}
Our hybrid system of interest consists of a NR classical
canonical subsystem and a NR quantized subsystem. To model their
coupled dynamics we start from the naive combination \eqref{Ale}.
In addition to the Dirac and Poisson brackets, there are mandatory
decoherence and diffusion terms that will necessitate the postulation of
a Riemann metric on the phase space manifold (or on its submanifold).
The resulting irreversible dynamics obtain the form of the hybrid master
equation (HME) which is the combination of the classical Fokker--Planck
and the quantum Lindblad equations (sec. \ref{subsec_HME}). 
This irreversible dynamics is equivalent  with the coupled stochastic
processes in the classical phase space and the Hilbert space, respectively,
and represented by a couple of hybrid stochastic differential equations (HSDEs)
in sec. \ref{subsec_SDE}. In physics, the special case is of interest, when
the classical coordinates are coupled to the quantum subsystem
but the classical momenta aren't (sec. \ref{subsec_coord}).
The material presented here is based primarily on refs. 
\cite{oppenheim2022two,oppenheim2023objective,diosi2023hybrid,diosi2023erratum}, 
and deduced basically from \cite{diosi2023hybrid} (cf. Appendix \ref{A}),
improved by the Riemann metric interpretation of  
the decoherence and diffusion kernels. 
It is important that we treat the HME and HSDE formalisms as equivalent, 
both have their own conceptual universality. 

\subsection{Hybrid master equation}\label{subsec_HME}
Let $\Ho(x)\equiv\Ho(q,p)$ be a our hybrid Hamiltonian where the 
classical subsystem is canonical.
The first $N$ canonical variables $\{x^n; n=1,\dots,N\}$ are the coordinates
and the second $N$ ones $\{x^n; n=N+1,\dots,2N\}$ are the momenta:
\beq
x^n=\left\{\begin{array}{cl}q_n; &n=1,2,\dots N\\
                                                     p^n; &n=N+1,N+2,\dots,2N.
                      \end{array}\right.
\eeq
The HME of the hybrid density $\ro(q,p)=\ro(x)$ takes this form:
\beq\label{HME}
\frac{d\ro}{dt}=-\frac{i}{\hbar}[\Ho,\ro]+\Herm\{\Ho,\ro\}+\bD\ro
                           \equiv\{\Ho,\ro\}_\mathrm{A}+\bD\ro,
\eeq
where $\{~,~\}_\mathrm{A}$ is the Aleksandrov hybrid bracket,
$\bD$ is the superoperator of decoherence and diffusion (D\&D). 
The notation of $x$-dependences of $\Ho,\ro,\bD$ are spared. 
The classical canonical Poisson bracket is defined by
\beq\label{Pois}
 \{\Ao,\Bo\}=\Ao\Dn\enm\Bo\Dm=(\Ao\enm\Bo\Dm)\Dn=(\Ao\Dn\enm\Bo)\Dm,
\eeq
where $\enm$ is the $2N\times2N$ symplectic matrix. We introduced the 
shorthand notation for partial derivatives like $\partial\Ao/\partial x^n=\Ao\Dn$
as well as Einstein's convention for summation of repeated indices.
If we define the canonical velocity operators 
\beq\label{v}
\vo^n=\{x^n,\Ho\}=-\enm\Ho\Dm
\eeq
then the Poisson bracket will have the useful equivalent expression: 
\beq\label{backa}
\Herm\{\Ho,\ro\}=\Herm\vo^n\ro\Dn.
\eeq

To construct the  canonical HME we impose a Riemann metric 
structure  in addition to the symplectic structure of the
phase space, via the arbitrary choice of the $2N\times2N$
covariant metric tensor  $\ga_{nm}(x)$. 
The D\&D terms are the following:
\bea\label{D}
\bD\ro&=&-\frac{\ga_{nm}}{8}[\vo^n,[\vo^m,\ro]]+\tfrac12\left(\ga^{nm}\ro\right)_{,nm}\nn\\
             &\equiv&\bD_\mathrm{Q}\ro+\bD_\mathrm{C}\ro,
\eea
where we assume that the velocities $\vo^n(x)$ are linearly independent
operators, also independent from any c-number functions. 
That is, we assume the equation
\beq\label{constr}
\la_n(x)\vo^n(x)=\varphi(x)
\eeq 
is satisfied only for vanishing $\la_n$ and $\varphi$.   

\subsection{Hybrid stochastic differential equations}\label{subsec_SDE}
The canonical HME \eqref{HME} with D\&D \eqref{D} 
is equivalent to two coupled stochastic processes, one for the
diffusion of  the pure state $\Po_t\equiv|\Psi_t\rangle\langle\Psi_t|$ 
in the Hilbert space, the other one for the diffusion of $x_t$ in the phase space,
meaning in fact the statistical interpretation of the HME.
Also called stochastic \emph{unraveling} of the HME, the processes are defined
by the couple of HSDEs:
\bea
\!\!\!\!\!\!\frac{d\Po}{dt}\!\!\!&=&\!\!\!-\frac{i}{\hbar}[\Ho(\!x\!),\!\Po]
\!+\!\bD_\mathrm{Q}(\!x\!)\Po
 \!+\!\Herm(\vo^n(\!x\!)\!-\!\langle\vo^n(\!x\!)\rangle)\Po w_n(\!x\!)\nn\\
\label{dP}\\
\label{dx}
\!\!\!\frac{d x^n}{dt}\!\!\!&=&\langle\vo^n(x)\rangle+w^n(x)
\eea
where $\langle\vo^n(x)\rangle=\tr(\vo^n(x)\Po)$. Both SDEs are driven
by the same white-noise $w_n=\ga_{nm}w^m$ whose correlations are
determined by the metric: 
\bea\label{ww}
\Mean w^n(x,t)w^m(x,\tau)&=&\ga^{nm}(x)\delta(t-\tau)\nn\\
\Mean w_n(x,t)w_m(x,\tau)&=&\ga_{nm}(x)\delta(t-\tau)\nn\\
\Mean w^n(x,t)w_m(x,\tau)&=&\delta^n_m\delta(t-\tau).
\eea 
The symbol $\Mean$ stands for the stochastic mean.

In this formalism of the hybrid dynamics
the backaction follows from the monitoring-plus-feedback mechanism.
The eq. \eqref{dP} coincides with the stochastic master equation of
time-continuous simultaneous quantum measurements ---monitoring--- 
of the observables $\vo^n$.
The measured signal $\langle\vo^n\rangle+w^n$ will then control a
feedback in the equation of motion \eqref{dx} of the classical phase 
space variables $x^n$. Note that this SDE can be written as 
\beq
\frac{dx^n}{dt}=\{x^n,\langle\Ho(x)\rangle\}+w^n,
\eeq
which is the mean-field (semiclassical) backaction plus our mandatory
white-noise. Observe that unlike white-noises, the phase-space 
coordinates $x^n(t)$ are continuous functions, containing the integrals
of the white-noises $w^n(t)$. 
The path in phase space is  a (generalized) Wiener process.
 
\subsection{Coordinate coupling}\label{subsec_coord}
The D\&D terms  \eqref{D} correspond
to the minimum noise 
dynamics if the $2N$ velocities $\vo^n(x)$ are independent 
operator fields on the phase space. However, they are not so in many
concrete hybrid systems. Suppose $K$ is the maximum number  of independent
constraints \eqref{constr}:
\beq
\la^a_n(x)\vo^n(x)=\varphi^a(x),~~~(a=1,2,\dots,K)
\eeq
with $K$ linear independent vector fields $\la^a_n\neq0$.  
Then  we can always find a coordinate transformation $x^n\Rightarrow f^n(x)$
such that the first $2N-K$ velocities $\vo^n$ become independent operators 
and the rest of them are c-numbers: $\vo^n=v^n\Io$ for $n\rangle2N-K$. 
Then the minimum noise D\&D corresponds to the same structure 
\eqref{D} but the indices run from $1$ to $2N-K$. The $(2N-K)\times(2N-K)$
metric tensor $\ga_{nm}$ defines a Riemann structure on the first $2N-K$ 
coordinates while it depends parametrically on the rest of them.

An important special case is coordinate-coupling when  $\partial\Ho/\partial q^n$
are independent operators but $\partial\Ho/\partial p^n$ are zeros or c-number
functions.
We impose the Riemann metric structure on the subspace
of canonical coordinates only. The $N\times N$ metric tensor 
$\ga_{nm}(q,p)$ will be the metric for the coordinates $q$ still it
may parametrically depend on the momenta $p$ as well.
With the hybrid part of momentum velocity operators 
\beq\label{vcoord}
\vo^n=-\frac{\der\HoCQ}{\der q_n}, 
\eeq
the D\&D terms take this form:
\beq\label{Dcoord}
\bD\ro=-\frac{\ga_{nm}}{8}
\left[\frac{\der\HoCQ}{\der q_n},\left[\frac{\der\HoCQ}{\der q_m},\ro\right]\right]
+\frac12\frac{\der^2\left(\ga^{nm}\ro\right)}{\der p^n\der p^m}~. 
\eeq
As we see, momentum velocity operators $\vo^n$ are actors of decoherence 
and classical momenta $p^n$ are subjects of diffusion.

The HSDEs (\ref{dx},\ref{dP}) of the equivalent  stochastic processes become
the following:
\bea
\label{dPcoord}
\frac{d\Po}{dt}&=&-\frac{i}{\hbar}[\Ho(q,p),\Po]+\bD(q,p)\Po+\nn\\
&&+\Herm\left(\vo^n(q,p)-\langle\vo^n(q,p)\rangle\right)\Po w_n(q,p)\\
\label{dqcoord}
\frac{d q_n}{dt}&=&\frac{\der\langle\Ho(q,p)\rangle}{\der p^n}\\
\label{dpcoord}
\frac{d p^n}{dt}&=& -\frac{\der\langle\Ho(q,p)\rangle}{\der q_n}+w^n(q,p).
\eea
Like in eq. \eqref{ww}, the noise $w^n=\ga^{nm}w_m$ satisfies
\bea
\label{wwcoord}
\Mean w^n(q,p,t)w^m(q,p,\tau)&=&\ga^{nm}(q,p)\delta(t-\tau)\nn\\
\Mean w_n(q,p,t)w_m(q,p,\tau)&=&\ga_{nm}(q,p)\delta(t-\tau)\nn\\
\Mean w^n(q,p,t)w_m(q,p,\tau)&=&\delta^n_m\delta(t-\tau).
\eea
This is the minimum-noise D\&D term of general coordinate coupling
provided the derivatives $\der\Ho/\der q_n$ are $N$ independent operators.

\section{Locality condition of relativistic continuum dynamics}\label{sec_rel} 
Let us consider the Markovian dynamics $d\rho/dt=L\rho$
where $\rho$ is classical, quantum, or hybrid state and
$L$ is the generator of time evolution respectively of
Fokker--Planck, Lindblad, or hybrid field dynamics. 
For relativistic invariance, 
$L$ must be the zeroth component of a four-vector. 
This condition on $L$ is, however, not sufficient \cite{diosi2022isthere}. 
It must be the spatial integral of the generator density $\Lc(\rv)$: 
\beq
L=\int\Lc(\rv)d\rv
\eeq
and $\Lc(\rv)$ must satisfy the locality condition 
\beq
[\Lc(\rv),\Lc(\sv)]=0.
\eeq
Then, given the state on the hypersurface $\si_1$, it maps to another
hypersurface as follows:
\beq
\rho(\si_2)=\exp\left(\int_{\si_2\succ(t,\rv)\succ\si_1}\hskip-40pt\Lc(\rv)d\rv dt\right)\rho(\si_1).
\eeq 
Without the locality condition, this relationship does not exist and 
we miss the map between states on two different hypersurfaces.
Of course, the map between Lorentz frames is also impossible.

In standard relativistic field theories, classical or quantum, the generator field
reads $\Lc=\{\Hc,~\}$ or  $\Lc=-(i/\hbar)[\Hco,~]$, respectively, and is local
since the Hamiltonian densities $\Hc,\Hco$ are local. Locality of the
generator $\Lc$ survives in effective field theories. If, however, 
the effective theory contains diffusion (or decoherence) then we face
difficulties. To retain locality of the generator $\Lc$ the diffusion (decoherence)
kernel must be local, i.e., proportional to $\delta(\rv-\sv)$ and 
then, unfortunately,  the theory yields infinities. 
Take, for instance, the Fokker--Planck equation of a scalar field
with the local diffusion kernel $\ga\delta(\rv-\sv)$.
It yields an infinite rate kinetic energy production at each point $\rv$. 
It is not known whether relativistic Fokker--Planck field equations
are renormalizable  or aren't. The same concern applies to the
Lindblad and hybrid dynamics.

\section{On special relativistic hybrid field dynamics}\label{sec_specrel}
We test the NR hybrid classical-quantum theory 
(sec. \ref{sec_nonrel}) in coordinate coupling (sec. \ref{subsec_coord}) of special 
relativistic fields. The coordinates and momenta become functions $q(\rv),p(\rv)$, 
and the discrete labels $n,m$ become the continuous spatial vectors $\rv,\sv$, respectively. 
Sums over indices become spatial integrals, Kronecker deltas become Dirac deltas,
derivations like e.g. $\der/\der q_n$ become functional derivations 
$\delta/\delta q(\rv)$.

Consider the coupling of the free classical scalar field $q(\rv)$ (with canonical momentum
$p(\rv)$) to the free quantized boson field $\Phio(\rv)$ (with canonical momentum $\hat\pi(\rv)$): 
\beq
\HoCQ[q]=\kappa\int q(\rv)\Phio(\rv)d\rv.
\eeq
This coupling is independent of the  classical canonical momentum $p(\rv)$ and we
can apply the eq. \eqref{Dcoord} with 
\beq
\frac{\delta\HoCQ}{\delta q(\rv)}=-\kappa\Phio(\rv). 
\eeq
The D\&D terms  depend on the metric which can in general  
be a functional kernel $\ga_{[q,q']}$. At the same time, we should 
damp remote correlation  in decoherence as well as in diffusion. 
The metric must have a spatial damping factor. In the simplest case,
we choose  a flat metric $\ga_{\rv\rv'}$ without the functional dependencies.
The covariant and contravariant kernels are inverses of each other:
\beq
\int\ga_{\rv\sv'}\ga^{\sv'\sv}d\sv'=\delta(\rv-\sv).
\eeq
Then the D\&D terms \eqref{Dcoord} take the following form:
\bea\label{Dqphi}
\bD\ro=&-&\frac{\kappa^2}{8}
\int\int\ga_{\rv\sv}\left[\Phio(\rv),\left[\Phio(\sv),\ro\right]\right]d\rv d\sv\nn\\
&+&\frac12\int\int\ga^{\rv\sv}\frac{\delta^2\left(\ro\right)}{\delta p(\rv)\delta p(\sv)} 
d\rv d\sv.
\eea

Both D\&D violate the special relativistic invariance 
unless the kernel itself is invariant. It is easy to
ensure Galilean invariance if $\ga_{\rv\sv}$ is a function of $\vert\rv-\sv\vert$.
The only kernels that ensure relativistic invariance are the singular local ones: 
\beq
\ga_{\rv\sv}=\ga\delta(\rv-\sv),~~~~\ga^{\rv\sv}=\ga^{-1}\delta(\rv-\sv).
\eeq
But they lead to untractable divergences of the  kinetic energy density 
$\mathcal{K}=\tfrac12(\hat\pi^2+p^2)$: 
\beq
\frac{d\mathcal{K}(\rv)}{dt}=
\frac12\bD_\mathrm{Q}\dg\pi^2(\rv)+\frac12\bD_\mathrm{C}\dg p^2(\rv)
=\left(\frac{\ga}{4\hbar^2}+\frac{1}{\ga}\right)\delta(\mathbf{0}).
\eeq
The D\&D terms \eqref{Dqphi} yield infinite rate of
heating at each location in the quantized bosonic as well as in the 
classical scalar field subsystems. 
Allowing functional dependence of the metric does not help since the
relativistic invariance of spatial damping requires the presence of the
spatial $\delta$ function singularity.  

These divergences are different from the  usual divergences in 
relativistic field theory. Either we invent their renormalization,
if it is possible at all, or we are  losing special relativistic invariance,
and we are left with the NR hybrid calculus.

\section{On hybrid general relativity}\label{sec_genrel}
Instead of full quantum-gravity, it were of great simplification if we could 
keep the space-time classical. 
Accordingly, we take a chance to extend the NR 
hybrid dynamics of Sec. \ref{sec_nonrel} for coupling between classical
canonical form of general relativity and quantized relativistic matter. 
In the canonical form of Einstein's general relativity, 
$(3+1)$-dimensional diffeomorphism invariance 
is encoded by the combination of $3$-dimensional spatial
diffeomorphism (sDM) invariance  and time-reparametrization (tRP) invariance. 
Following refs. \cite{oppenheim2023gravitationally,oppenheim2023postquantum},
we build up the formal sDM and tRP invariant hybrid equations
(sec. \ref{subsec_genrelHMEHSDE}).  
We are \emph{going to the wall} to ensure both these invariances 
but that remains a problem ( \ref{subsec_genrelDD}). 

\subsection{Equivalent formalisms: HME and HSDE}\label{subsec_genrelHMEHSDE}
The canonical coordinates are the configurations of the $3\times3$ 
metric tensor field $g_{ik}(\rv)$, satisfying the canonical commutation
relationship with the canonical momenta $\pi^{ik}(\rv)$:
\beq
\{g_{ij}(\rv),\pi^{kl}(\sv)\}=\delta_{ij}^{kl}\delta(\rv,\sv),
\eeq
where $\delta_{ij}^{kl}=\frac12(\delta_i^k\delta_j^l+\delta_i^l\delta_j^k)$
and we use the covariant delta function 
\beq
\delta(\rv,\sv)=\frac{1}{\sqrt{g(\rv)}}\delta(\rv-\sv)~.
\eeq
where $g=\mathrm{det}g_{ij}$.
The covariant Poisson bracket is defined by
\beq
\{\Ao,\Bo\}=
\int \left(\frac{\delta\Ao}{\delta g_{ij}(\rv)} \frac{\delta\Bo}{\delta\pi^{ij}(\rv)}
                 -\frac{\delta\Ao}{\delta\pi^{ij}(\rv)} \frac{\delta\Bo}{\delta g_{ij}(\rv)}
\right)dV,
\eeq
where $dV=dV_\rv=\sqrt{g(\rv)}d\rv$. Through this section,  
the functional derivatives are the covariant ones, i.e., 
$1/\sqrtg$  times the common ones.

The hybrid Hamiltonian reads:
\beq\label{H_ADM}
\Ho[g,\pi;N,\vec{N}]=\HG[g,\pi;N,\vec{N}]+\HoM[g;N,\vec{N}],
\eeq
where $\HG[g,\pi;N,\vec{N}]$ is the classical Hamilton function of
gravity and $\HoM[g;N,\vec{N}]$ is the Hamiltonian of the quantized matter fields,
coupled only to $g_{ik}$ and not to $\pi^{ik}$. They depend on the
freely chosen lapse $N$ and shift $N_i$: 
\bea
\!\!\!\!\!\!\!\!\!\!\HG[g,\pi;N,\vec{N}]\!\!&=&\!\!\!\!
\int\!\! \left(N(\rv) \HcG(\rv)\!+\!N_i(\rv)\Pc^i_G(\rv)\right)\! dV,\\
\HoM[g;N,\vec{N}]\!\!&=&\!\!\!\!\int\!\!\left(N(\rv)\HcoM(\rv)+N_i\Pc^i_M(\rv)\right)dV.
\eea

$\HcG(\rv)$ and $\HcoM(\rv)$ are the Hamiltonian densities of gravity and matter, 
respectively, and $\Pc_G^i$ is the momentum density of gravity:
\beq
\Pc^i_G=-2\nabla_i\pi^{ij}(\rv),
\eeq
where $\nabla_{\!j}$ denotes covariant derivation. 
The gravity's Hamiltonian density reads:
\beq
\Hc_G=
\frac{16\pi G}{c^2}\frac{1}{g}\left(\pi^{ij}\pi_{ij}-\tfrac12(\pi^i_i)^2\right)-\frac{c^4}{16\pi G}R,
\eeq
with the scalar curvature $R$.
The matter's Hamiltonian $\hat{\Hc}_M(\rv)$ 
and momentum density $\Pc^i_M$ depend 
on the matter fields. Remember that they should not depend on $\pi^{ik}$ . 

In the hybrid Hamiltonian \eqref{H_ADM},
the lapse $N$ multiplies the Hamiltonian constraint, the shift $N_i$ multiplies
the diffeomorphism constraint which we impose on the hybrid state:
\bea
\label{CHam}
\left(\HcG(\rv)+\HcoM(\rv)\right)\ro[g,\pi]=0,\\
\label{Cmom}
\left(\Pc^i_G(\rv)+\hat{\Pc}^i_M(\rv)\right)\ro[g,\pi]=0.
\eea
These might ensure tRP and sDM invariances respectively.
The conditional phrase is of reason. If both gravity and matter
were quantized (or classical), then the above constraints  would guarantee
the said invariances under pure classical canonical  (or pure unitary)
dynamics. Their compatibility and applicability in hybrid dynamics are not yet clear.
Moreover, hybrid dynamics are not necessarily compatible with tRP and sDM
invariances, as we see below.

To construct the hybrid coupling and the D\&D terms, we need 
the momentum velocity operators \eqref{vcoord}:
\beq
\vo^{ik}(\rv)=-\frac{\delta\Ho_M}{\delta g_{ik}(\rv)}
                       =-N(\rv)\!\left(\!\frac{\der\hat{\Hc}_M(\rv)}{\der g_{ik}(\rv)}
                             +\frac12 g^{ik}(\rv)\hat{\Hc}_M(\rv)\!\right).
\eeq
The HME \eqref{HME} of the state $\ro[g,\pi]$ takes this form:
\beq\label{HME_ADM}
\frac{d\ro}{dt}=-\frac{i}{\hbar}[\Ho_M,\ro]+\{H_G,\ro\}
-\Herm\int
\vo^{ik}\frac{\delta\ro}{\delta\pi^{ik}}dV+\bD\ro\\
\eeq
While the hybrid Hamiltonian parts are unique, 
the D\&D term $\bD\ro$ is not, its consistent choice is nontrivial,
see sec. \ref{subsec_genrelDD}.  

The HME \eqref{HME_ADM} has its alternative stochastic representation
in terms of SHDEs.
We apply the eqs. (\ref{dqcoord}-\ref{wwcoord}):
\bea
\label{dP_ADM}
\!\!\!\frac{d\Po}{dt}\!\!\!&=&\!\!-\frac{i}{\hbar}[\HoM,\Po]\!+\!\bD_\mathrm{Q}\Po
\!+\!\!\Herm\!\!\int\!\!\left(\vo^{ij}\!\!-\!\!\langle\vo^{ij}\rangle\right)\!\!\Po w_{ij}dV\\
\label{dg_ADM}
\!\!\!\frac{d g_{ij}}{dt}\!&=&\frac{\delta\HG}{\delta\pi^{ij}}\\
\label{dpi_ADM}
\!\!\!\frac{d\pi^{ij}}{dt}\!&=&-\frac{\delta\HG}{\delta g_{ij}}
                                           +\langle\vo^{ij}\rangle+w^{ij}
\eea
where  the noises satisfy
\bea\label{ww_ADM}
\Mean w^{ij}(\rv,t)w^{kl}(\sv,\tau)&=&\ga^{ij|kl}(\rv|\sv)\delta(t-\tau)\nn\\
\Mean w_{ij}(\rv,t)w_{kl}(\sv,\tau)&=&\ga_{ij|kl}(\rv|\sv)\delta(t-\tau)\nn\\
\Mean w^{ij}(\rv,t)w_{kl}(\sv,\tau)&=&\delta^{ij}_{kl}\delta(t-\tau),
\eea
and $\bD_\mathrm{Q}$ will be discussed in sec. \ref{subsec_genrelDD}.

As we said in sec. \ref{subsec_SDE}, 
the eq. \eqref{dP_ADM} corresponds to the quantum
monitoring of the velocity operators $\vo^{ij}=-\delta\HoM/\delta g_{ij}$ 
and the noisy measured signal $\langle\vo^{ij}\rangle+w^{ij}$ is fed back
on the rhs of the eq. \eqref{dpi_ADM} of $d\pi^{ij}/dt$.

\subsection{The decoherence-diffusion kernels}\label{subsec_genrelDD}
Recall that the hybrid dynamics (sec. \ref{subsec_coord}) assumes a certain 
metric on the space
of canonical coordinates, which is a functional metric 
on the function space of $3\times3$ metric tensor fields $g_{ij}(\rv)$.  
We restrict ourselves to the metrics
\beq
(dg)^2=\int\int\ga^{ij|kl}(\rv|\sv)dg_{ij}(\rv)dg_{kl}(\sv)dV_\rv dV_\sv,
\eeq
where the functional metric tensor $\ga$ contains explicit
coordinate dependence on $(\rv,\sv)$ to damp remote correlations,
also it may depend on $g_{ij}(\rv)$ and $g_{kl}(\sv)$ (meaning nonflat
functional geometry). Accordingly, the D\&D terms take this form:
\bea
\label{DQG}
\bD_\mathrm{Q}\!\!\!&=&\!\!\!-\frac{1}{8}\!\int\!\!\!\int\!\!
\ga^{-1}_{ij|kl}(\rv|\sv)\left[\vo^{ij}(\rv),[\vo^{kl}(\sv),\ro]\right]dV_\rv dV_\sv\\
\label{DCG}
\bD_\mathrm{C}\!\!\!&=&\frac12\int\!\!\!\int\!
\frac{\delta^2\left(\ga^{ij|kl}(\rv|\sv)\ro\right)}{\delta\pi^{ij}(\rv)\delta\pi^{kl}(\sv)}
      dV_\rv dV_\sv,        
\eea
where the covariant and contravariant metrics satisfy the functional
relationship
\beq\label{gacovgacontr}
\int\int\ga_{ij|k'l'}(\rv|\sv')\ga^{k'l'|kl}(\sv'|\sv)dV_{\sv'}=\delta_{ij}^{kl}\delta(\rv,\sv)~.
\eeq
It is instructive to consider the simple special case when the kernels are local.
Then their structure is perfectly determined by covariance:
\bea
\label{gcovGloc}
\ga_{ij|kl}(\rv|\sv)&=&\frac{\ga(R)}{N(\rv)}G^{(\al)}_{ij|kl}(\rv)\delta(\rv,\sv)\\
      G^{(\al)}_{ij|kl}&=&\tfrac12 g_{ik}g_{jl}+\tfrac12 g_{il}g_{jk}+\al g_{ij}g_{kl}\nn\\
\label{gconGloc}
\ga^{ij|kl}(\rv|\sv)&=&\frac{N(\rv)}{\ga(R)}G_{(\be)}^{ij|kl}(\rv)\delta(\rv,\sv)\\
G_{(\be)}^{ij|kl}&=&\tfrac12 g^{ik}g^{jl}+\tfrac12 g^{il}g^{jk}+\be g^{ij}g^{kl}.\nn
\eea
These kernels are positive if $\al,\be\rangle-1/3$.
If $3\al\be+\al+\be=0$ then the kernels are each other's inverses as they should,
according to eq. \eqref{gacovgacontr}.

With the above kernels, unfortunately,  both the D\&D terms  
in eqs. \eqref{DQG} and \eqref{DCG}, resp., become
divergent because of the $\delta$-functions, just like in sec. \ref{sec_specrel}.
However, a rescue procedure seems to be on offer.

We could try the sDM invariant regularization.
For instance, we replace the $\delta(\rv,\sv)$ in the decoherence kernel
\eqref{gcovGloc} by
\beq
\mathcal{N}_\eps(\rv,\sv)\exp\left(-\frac{\ell^2(\rv,\sv)}{2\eps}\right)
\eeq
where  $\mathcal{N}_\eps(\rv,\sv)$ is for normalization,
$\ell(\rv,\sv)$ is the geodesic distance between $\rv$ and $\sv$,
and $\eps$ is the small parameter to go to $+0$.
To keep covariance,  the index factor, too, should go nonlocal:
\bea
G^{(\al)}_{ij|kl}(\rv|\sv)
\!\!&=&\!\!\tfrac12 P^{j'}_j \bar{P}^{k'}_k g_{ik'}(\rv)g_{j'l}(\sv)
     \!+\!\tfrac12  P^{j'}_j \bar{P}^{l'}_l g_{il'}(\rv)g_{kj'}(\sv)\nn\\
&&+\al g_{ij}(\rv)g_{kl}(\sv).
\eea
Here $P^i_j$ is geodesic parallel transport of covariant vectors from $\sv$ to $\rv$
and $\bar{P}^i_j$ is the same from $\rv$ to $\sv$.

So far so good. The problem is the factor $1/N(\rv)$ which ensures 
the tRP  invariance. We should keep it but we cannot.
It cannot be split for the two locations $\rv$ and $\sv$.
The same problem would come along with the factor $N(\rv)$  if we regularized
the decoherence kernel \eqref{gconGloc} first.

The lesson goes beyond the example. Any nonlocal generalization
of the kernels will necessarily violate the tRP invariance. Local kernels,
on the other hand, generate divergences whose removal may or may
not be possible. Hence, for the time being,  a compromise seems inevitable.
We give up tRP invariance and retain sDM invariance that allows regular
nonlocal kernels. Just losing tRP invariance means losing relativistic
invariance.  We are left with NR slow motions in a distinguished frame:
sDM is pointless. Also the space-time must be nearly flat. That's 
the Newtonian limit. 

\section{Newtonian hybrid classical-quantum gravity}\label{sec_DP}
When recapitulating the results of 
refs. \cite{tilloydiosi2016sourcing,tilloydiosi2017},
we use a particular approach.  These works used the NR
HSDE representation of hybrid dynamics. Not for deduction but 
for comparison, we guide our derivation by the HSDEs (\ref{dP_ADM}-\ref{dpi_ADM}) 
that promised general relativistic postquantum gravity in sec. \ref{sec_genrel}. 
We present  the HSDEs of Newtonian hybrid theory first. 

What is the closest NR dynamics to the HSDEs (\ref{dP_ADM}-\ref{ww_ADM})?
The matter Hamiltonian with the hybrid coupling reads
\beq
\HoM[\Phi]=\Ho_0+\int\muo\Phi dV
\eeq
where $\Phi$ is the Newton potential and $\muo$ is the NR quantum field of mass density. 
The  quantum monitoring  of $\vo^{ij}$ corresponds to the quantum monitoring of 
$\hat{\mu}(\rv)$  since the nonrelativistic limit of $\vo^{ij}$ is $\propto\muo$.
Hence the NR counterpart of the SDE \eqref{dP_ADM} 
\beq
\label{dP_DP}
\frac{d\Po}{dt}\!=\!-\frac{i}{\hbar}[\HoM[\Phi],\Po]\!+\!\bD_\mathrm{Q}\Po
\!+\!\frac{1}{\hbar}\Herm\!\!\int\!\!\left(\muo\!-\!\langle\muo\rangle\right)\!\Po w dV,
\eeq
with
\beq\label{DQ_DP}
\bD_\mathrm{Q}\Po
=-\frac{1}{8\hbar^2}\int\int\ga_{\rv\sv}\left[\muo(\rv),[\muo(\sv),\Po]\right]d\rv d\sv.
\eeq   
The measurement signal is of the standard form 
\beq\label{signal}
\langle\muo\rangle+\wt,
\eeq
where $\wt(\rv,t)=\int\ga^{\rv\sv}w(\sv,t)d\sv$. 
The covariant and  contravariant components ($w,\wt$) of the same noise
satisfy 
\bea
\Mean w(\rv,t) w(\sv,\tau)&=&\ga_{\rv\sv}\delta(t-\tau)\nn\\
\Mean \wt(\rv,t)\wt(\sv,\tau]&=&\hbar^2\ga_{\rv\sv}\delta(t-\tau)\nn\\
\Mean w(\rv,t)\wt(\sv,\tau]&=&\hbar\delta(\rv-\sv)\delta(t-\tau).
\eea
Since gravity has no self-dynamics, $\HG=0$, the backaction 
(\ref{dg_ADM},\ref{dpi_ADM}) reduces to the Poisson equation sourced
by the signal \eqref{signal} and we can solve it:
\bea
\Phi(\rv,t)&=&\frac{4\pi G}{\nabla^2}\left(\langle\muo(\rv)\rangle_t+\wt(\rv,t)\right)\nn\\
                   &\equiv&\Phimf(\rv,t)+\delta\Phi(\rv,t).
\eea
The deterministic term $\Phimf$ is the mean-field (semiclassical) part,
the stochastic term is a white-noise of correlation
\beq
\Mean \delta\Phi(\rv,t)\delta\Phi(\sv,\tau)
=\frac{4\pi G}{\nabla^2_\rv}\frac{4\pi G}{\nabla^2_\sv}\hbar^2\ga_{\rv\sv}\delta(t-\tau).
\eeq

When $\Phi$ is fed back in eq. \eqref{dP_DP},  the Hamiltonian $\HoM[\Phi]$ 
generates the Newtonian pair potential 
\beq
\VoG=-\frac{G}{2}\int\int\frac{\muo(\rv)\muo(\sv)}{|\rv-\sv|}d\rv d\sv.
\eeq
Unlike the general relativistic $\HoM[g]$, where $g$ is a Wiener process, 
$\Phi$ is not, it is the time-derivative of a Wiener process. 
The feedback of the white-noise term in $\HoM[\Phi]$, proportional to  $\delta\Phi$,  
will contribute to a new decoherence term:
\beq\label{DQ_DPFB}
\bD_\mathrm{Q}^\mathrm{fb}\Po
=-\frac{1}{2\hbar^2}\int\!\!\!\int
\!\!\left(\frac{4\pi G}{\nabla^2_\rv}\frac{4\pi G}{\nabla^2_\sv}\ga^{\rv\sv}\right)
         \! \left[\muo(\rv),[\muo(\sv),\Po]\right]\!d\rv d\sv.
\eeq
This backaction makes a remarkable difference  
compared to the general relativistic case in sec. \ref{subsec_genrelDD}.
The ambiguity of the D\&D kernels can be removed by the principle
of least decoherence.
Since $\ga^{\rv\sv}=\ga_{\rv\sv}^{-1}$, the total decoherence 
$\bD_\mathrm{Q}+\bD_\mathrm{Q}^\mathrm{fb}$ possesses a minimum
when
\bea
\ga_{\rv\sv}&=&\frac{2\hbar G}{|\rv-\sv|},\nn\\
\ga^{\rv\sv}&=&-\frac{1}{8\pi\hbar G}\nabla^2\delta(\rv-\sv).
\eea
Accordingly, the least decoherence reads
\beq\label{DQ_DPmin}
\bD_\mathrm{Q}^\mathrm{DP}=
-\frac{G}{2\hbar}\int\int\frac{\left[\muo(\rv),[\muo(\sv),\Po]\right]d\rv d\sv}
{|\rv-\sv|}
\eeq
and the correlation of the gravitational fluctuations become
\beq\label{dPhidPhi}
\Mean\delta\Phi(\rv,t)\delta\Phi(\sv,\tau)=\frac{\hbar G/2}{|\rv-\sv|}\delta(t-\tau).
\eeq

We obtain  the HSDEs of the Newtonian NR postquantum gravity:
\bea
\label{dP_DPFBmin}
\!\!\!\!\!\frac{d\Po}{dt}\!\!&=&\!\!
-\frac{i}{\hbar}[\Ho_0\!+\!\!\VoG,\!\Po]\!+\!\!\bD_\mathrm{Q}^\mathrm{DP}\!\Po
\!+\!\Herm\frac{1\!\!+\!i}{\hbar}\!\!\!\int\!\!\!\left(\!\muo\!-\!\!\langle\muo\rangle\!\right)\!\Po w dV\\
\label{dPhi_DPFBmin}
\!\!\!\!\!\Phi&=&\frac{4\pi G}{\nabla^2}\langle\muo\rangle-\frac12 w
                         =\Phimf-\frac12 w,
\eea
where $\Phimf$ is the mean-field (semiclassical) Newton potential, and
\beq
\label{ww_DP}
\Mean w(\rv,t)w(\sv,\tau)=\frac{2\hbar G}{|\rv-\sv|}\delta(t-\tau).
\eeq
 
For point-like particles the theory is divergent, predicts kinetic energy increase
at infinite rate. Therefore $\muo(\rv)$ must be smoothened by a short length cutoff
parameter, the only free parameter of the theory 
(see \cite{donadi2021underground} for its experimental limit).

Observe that due to the simple structure of the Newtonian postquantum
dynamics the reduced dynamics of the quantized matter is autonomous.
Take the stochastic mean of both sides of eq. (\ref{dP_DPFBmin}) then
the following Lindblad master equation is obtained for 
$\ro_\mathrm{Q}=\Mean\Po$: 
\beq
\frac{d\ro_\mathrm{Q}}{dt}=
-\frac{i}{\hbar}[\Ho_0+\VoG,\ro_\mathrm{Q}]
-\frac{G}{2\hbar}\int\!\!\!\!\int[\muo(\rv),[\muo(\sv),\ro_\mathrm{Q}]]\frac{d\rv d\sv}{|\rv-\sv|}.
\eeq
The  full HME, equivalent to the HSDE formalism  (\ref{dP_DPFBmin}-\ref{ww_DP}),
 is derived in Appendix \ref{B}. 

\section{Remarks, Conclusions}\label{sec_remarks}
The issues of hybrid dynamics relativistic extensions that  secs.
\ref{sec_specrel} and \ref{sec_genrel} claim are unsolved, 
were carefully discussed by the authors of refs. 
\cite{oppenheim2023gravitationally,oppenheim2023postquantum},
highlighting some perspectives towards solutions. 
These are assessed with certain reservations in ref. \cite{tilloy2024general}. 
We add that the literature offers no support for hybrid constraints, 
little or no support for renormalizability of relativistic effective field theories 
let them be classical, quantum, or hybrid.
Towards  fixing infinities predicted by relativistic Lindblad and Fokker--Planck
equations,  conclusive research is missing even for the 
the simple special relativistic D\&D in sec. \ref{sec_specrel}.



Some additional details about  
the nonrelativistic `postquantum' theory (sec. \ref{sec_DP}) are to be recalled.
It all started in foundations (reviewed in \cite{bassi2013models,bassi2023collapse}), 
with a gravity-related nonrelativistic model of the
quantum-classical transition \cite{diosi1989} and a naive formalism
of relativistic monitoring-plus-feedback \cite{diosi1990}. 
Recognizing the difficulties of relativistic monitoring, 
only the Newtonian limit of monitoring-plus-feedback was briefly presented. 
Much later, the concept of postquantum gravity, called  a `conceptually healthier 
semiclassical theory', was stated literally in \cite{tilloydiosi2016sourcing}: 
monitoring the quantized energy-momentum
tensor $\hat T_{ab}$ and its measured value fed back into the
Einstein equation of classical general relativity. 
After two and a half decades,  this work and its followup \cite{tilloydiosi2017} 
must still have adhered to the Newtonian limit.
The reason has remained the same: missing theory of
relativistic monitoring. The concrete technical obstacles are 
the D\&D kernels that must be time-local for Markovianity. 
If the suitably covariant kernels exist at all, they generate divergences
whose treatment is unknown. 
Without these difficulties, the monitoring-plus-feedback
form (equivalent to the hybrid master equation form) of postquantum general relativity
would have been a straightforward step. Vice versa, if the hybrid master
equation form of postquantum gravity got rid of its difficulties with
the D\&D kernels, it would contain a modul of relativistic quantum monitoring. 
This matches with the assessment in ref. \cite{tilloy2024general}.
 
The pending issues of the recent proposal 
\cite{oppenheim2023gravitationally,oppenheim2023postquantum}
of postquantum gravity
are the known old difficulties that have been hindering the relativistic extension of the
Newtonian `forerunner' 
\cite{diosi1989,diosi1990,tilloydiosi2016sourcing,tilloydiosi2017}.  
The difficulties are rooted in
difficulties of Lindblad as well as of  Fokker--Planck  dynamics of 
relativistic fields; both dynamics are obligatory parts of postquantum gravity. 
Although these issues might become fixed later, the contrary is equally
likely: relativity and Markovianity of decoherence (or diffusion)
may turn out to be just inconsistent \cite{diosi2022isthere}. 

In contrast to the relativistic postquantum gravity, the Newtonian precursor
\cite{diosi1989,tilloydiosi2016sourcing,tilloydiosi2017} 
is a consistent model with a single free parameter. The predicted violation of
the superposition principle and the presence of the tiny noise have been
looked for by various experiments reviewed e.g. in \cite{carlesso2022present}.
The model, also called the Di\'osi--Penrose model, is currently neither confirmed nor ruled out. 
For a conclusive test,  the quantum control of the test mass motional states must 
be improved. Even higher improvement will be requested in the proposed
nonrelativistic tests to rule out the classicality of gravity 
\cite{bose2017spin,marletto2017gravitationally,howl2021non}.
Such tests might or might not rule out unquantized gravity theories. 
As yet, this is at worst a period of grace for them.

The present author
expects that the hybrid of classical gravity and quantized matter is 
hiding more secrets already in the Newtonian limit, both in theory and
experiments. We should continue
to reveal them in the simple nonrelativistic realm 
before we would cross the bridge towards a certain postquantum general relativity.

\acknowledgments
I thank Isaac Layton, Jonathan Oppenheim, Andrea Russo and Antoine Tilloy for
illuminating discussions.
This research was funded by the Foundational Questions Institute and Fetzer Franklin
Fund, a donor-advised fund of the Silicon Valley Community Foundation 
(Grant No. FQXi-RFPCPW-2008), 
the National Research, Development and Innovation Office (Hungary)
``Frontline'' Research Excellence Program (Grant No. KKP133827),
and the John Templeton Foundation (Grant 62099).

\appendix
\section{Deduction of HME \eqref{HME}}
\label{A}
We show that our canonical HME \eqref{HME} with the D\&D term \eqref{D}
is the special case of the general diffusive HME 
\cite{oppenheim2022two,oppenheim2023objective,
diosi2023hybrid,diosi2023erratum}: 
\bea\label{genHME}
\frac{d\ro}{dt}&=&-i[\Ho,\ro]+2\Herm([\GCQb]^n_\al\Loa\ro)\Dn+\bD\ro\\
\label{genD}
\bD&=&\DQ_{\be\al}\Bigl(\Loa\ro\Lo^\be-\Herm\Lo^\be\Loa\ro\Bigr)
+\tfrac12 \left(\DC^{nm}\ro\right)_{nm}
\eea
where, compared to  eq. (36) in  \cite{diosi2023hybrid}, we assumed Hermitian Lindblad 
generators $\Loa$ and changed the upper/lower greek indices
for the lower/upper ones. This HME is valid for any classical subsystem, 
the classical coordinates $x$ are not necessarily canonical.
When the Lindblad generators $\Loa(x)$ are independent operators then
minimum noise is achieved if  the positive D\&D matrices $\DQ,\DC$,
resp., are constrained by the matrix of backaction $\GCQ$:
\beq\label{Opp}
\GCQ\frac{1}{\DQ}\GCQ\dg=\DC.
\eeq

Let us first identify the classical variables $x^n$ by our canonical ones.
Second, identify the Lindblad generators $\Loa$ by our velocity operators $\vo^n$,
the greek indices will become the latin ones accordingly.
Let us equate the backaction terms in \eqref{HME} and \eqref{genHME}: 
\beq
\Herm(\vo^n\ro)\Dn=-2\Herm([\GCQb]^n_{~m}\vo^n\ro)\Dn.
\eeq
They coincide if $[\GCQb]^n_m=-\tfrac12\delta^n_m$.
The D\&D terms \eqref{D} and \eqref{genD} coincide if 
$\DQ_{nm}=\tfrac14\ga_{nm}$ and $\DC^{nm}=\ga^{nm}$.
The said choices $\DC,\DQ$ and $\GCQ$
satisfy the general condition \eqref{Opp} of minimum noise.   

\section{Derivation of HME from HSDEs (\ref{dP_DPFBmin}-\ref{ww_DP})}
\label{B}
It is incorrect to  take the form $\ro[\Phi]$ for the hybrid state since 
$\Phi$ is a white noise. The correct form is $\ro_t[\chi]$, i.e., 
the configuration of classical gravity is represented by the Wiener process
$\chi$ defined by $\Phi=d\chi/dt$. We  define the hybrid density as follows:
\beq\label{rhochi}
\ro_t[\chi]=\Mean\Po_t\delta[\chi-\chi_t]
\eeq
The differentials of both sides read
\beq\label{drhochi}
d\ro_t[\chi]=\Mean\left(d\Po_t\delta[\chi-\chi_t]+\Po_t d\delta[\chi-\chi_t]
                                             +d\Po_t d\delta[\chi-\chi_t]\right)
\eeq
where the last term on the r.h.s. is the Ito correction to the Leibnitz rule.
According to Ito calculus, 
using the HSDEs (\ref{dP_DPFBmin},\ref{dPhi_DPFBmin}) and the white-noise
correlation (\ref{ww_DP}) yield
\bea
d\Po\!&=&-\frac{i}{\hbar}[\Ho_0+\VoG,\Po]dt+\bD_\mathrm{Q}^\mathrm{DP}\Po dt\nn\\
&&+\Herm\frac{1\!+\!i}{\hbar}\!\!\int\!\!\left(\muo(\rv)\!-\!\langle\muo(\rv)\rangle\right)\!\Po w(\rv,t) d\rv dt\\
d\delta[\chi\!-\!\chi_t]\!\!&=&
\!-\!\!\int\!\left(\Phimf(\rv)\!-\!\tfrac12 w(\rv,t)\right)\!
\frac{\delta}{\delta\chi(\rv)}\delta[\chi\!-\!\chi_t]d\rv dt\nn\\
&&\!+\!\frac14\!\!\int\!\!\!\!\!\int\!\!\!\frac{\hbar G}{|\rv\!-\!\sv|}
\frac{\delta^2}{\delta\chi(\rv)\delta\chi(\sv)}\delta[\chi\!-\!\chi_t]d\rv d\sv dt\\
d\Po d\delta[\chi\!-\!\chi_t]\!&=&\Herm(1\!+\!i)
\int\!\!\!\!\!\int\!\!\!\frac{G}{|\rv\!-\!\sv|}\!
\left(\muo(\sv)\!-\!\langle\muo(\sv)\rangle\right)\Po\times\nn\\
&&\times\frac{\delta}{\delta\chi(\rv)}\delta[\chi\!-\!\chi_t]d\rv d\sv dt.
\eea
Now we insert these three expressions into eq. (\ref{drhochi}),
set $w=0$ since $\Mean w=0$, and use the definition (\ref{rhochi}) of $\ro[\chi]$,
yielding, after dividing both sides by $dt$: 
\bea
\frac{d\ro[\chi]}{dt}=&&-\frac{i}{\hbar}[\Ho_0+\VoG,\ro[\chi]]+\bD_\mathrm{Q}^\mathrm{DP}\ro[\chi]\nn\\
&&-\!\!\int\!\Phimf(\rv)\!
\frac{\delta\ro[\chi]}{\delta\chi(\rv)}d\rv
\!\!+\frac14\int\!\!\!\!\!\int\!\!\!\frac{\hbar G}{|\rv\!-\!\sv|}\!
\frac{\delta^2\ro[\chi]}{\delta\chi(\rv)\delta\chi(\sv)}d\rv d\sv\nn\\
&&+\Herm(1\!+\!i)
\!\!\int\!\!\!\!\!\int\!\!\!\frac{G}{|\rv\!-\!\sv|}\!
\left(\muo(\sv)\!-\!\langle\muo(\sv)\rangle\right)
\frac{\delta\ro[\chi]}{\delta\chi(\rv)}d\rv d\sv.
\eea
The nonlinear terms on the r.h.s. cancel as they should and we
write the HME in the following form: 
\begin{widetext}
\beq\label{HME_DP}
\frac{d\ro}{dt}=
-\frac{i}{\hbar}[\Ho_0+\VoG,\ro]
+G\!\int\!\!\!\int\!
\left(-\frac{1}{2\hbar}[\muo(\rv),[\muo(\sv),\ro]]
          +\Herm(1\!+\!i)\muo(\rv)\frac{\delta\ro}{\delta\chi(\sv)}
          +\frac{\hbar}{4}\frac{\delta^2\ro}{\delta\chi(\rv)\delta\chi(\sv)}\right)
\frac{d\rv d\sv}{|\rv-\sv|}.
\eeq
\end{widetext}
The HME yields the mean-field (semiclassical) gravity: 
\beq
\Mean\Phi(\rv)=\tr\!\!\int\!\!\frac{d\chi(\rv)}{dt}\ro[\chi]d[\chi]=-G\!\!\int\!\!\frac{\langle\muo(\sv)\rangle}{|\rv-\sv|}d\sv=\Phimf(\rv),\nn
\eeq
as well as the `space-time'  diffusion (\ref{dPhidPhi}) where $\delta\Phi=d\chi/dt-\Phimf$.

\bibliography{diosi2024}{}
\end{document}